\documentclass{aastex}
\usepackage{emulateapj5}
\usepackage{apjfonts}

\newcommand{\swas}{{\it SWAS}}
\newcommand{\ci}{C~{\sc i}}
\newcommand{\hii}{H~{\sc ii}}
\newcommand{\thco}{$^{13}$CO}
\newcommand{\thcoo}{$^{13}$CO(1$\rightarrow$0)}
\newcommand{\thcot}{$^{13}$CO(2$\rightarrow$1)}
\newcommand{\thcof}{$^{13}$CO(5$\rightarrow$4)}
\newcommand{\kms}{km s$^{-1}$}
\newcommand{\cmsqr}{cm$^{-2}$}
\newcommand{\cmcube}{cm$^{-3}$}

\journalinfo{{\rm Published in} The Astrophysical Journal, 539:L137--L141,
             {\rm 2000 August 20}}
\slugcomment{Received 2000 May 1; accepted 2000 June 23; published
             2000 August 16}

\shorttitle{HOWE ET AL.}
\shortauthors{[\ci] AND $^{13}$CO(5$\rightarrow$4) EMISSION IN M17SW}

\begin{document}

\title{EXTENDED [\ci] AND $^{13}$CO(5$\rightarrow$4) EMISSION IN M17SW}

\author{J. E. Howe\altaffilmark{1},
M. L. N. Ashby\altaffilmark{2},
E. A. Bergin\altaffilmark{2},
G. Chin\altaffilmark{3},
N. R. Erickson\altaffilmark{1},
P. F. Goldsmith\altaffilmark{4},
M. Harwit\altaffilmark{5},
D. J. Hollenbach\altaffilmark{6},
M. J. Kaufman\altaffilmark{7},
S. C. Kleiner\altaffilmark{2},
D. G. Koch\altaffilmark{8},
D. A. Neufeld\altaffilmark{9},
B. M. Patten\altaffilmark{2},
R. Plume\altaffilmark{2},
R. Schieder\altaffilmark{10},
R. L. Snell\altaffilmark{1},
J. R. Stauffer\altaffilmark{2},
V. Tolls\altaffilmark{2},
Z. Wang\altaffilmark{2},
G. Winnewisser\altaffilmark{10},
Y. F. Zhang\altaffilmark{2}, AND
G. J. Melnick\altaffilmark{2}}

\altaffiltext{1}{Department of Astronomy, University of Massachusetts,
                 Amherst, MA  01003.}
\altaffiltext{2}{Harvard-Smithsonian Center for Astrophysics,
                 60 Garden Street, Cambridge, MA  02138.}
\altaffiltext{3}{NASA Goddard Spaceflight Center, Greenbelt, MD 20771.}
\altaffiltext{4}{National Astronomy and Ionosphere Center, Department
                 of Astronomy, Cornell University, Space Sciences Building,
                 Ithaca, NY  14853-6801.}
\altaffiltext{5}{511 H Street SW, Washington, DC  20024-2725; also
                 Cornell University.}
\altaffiltext{6}{NASA Ames Research Center, MS 245-3, Moffett Field, CA  94035.}
\altaffiltext{7}{Department of Physics, San Jose State University,
                 One Washington Square, San Jose, CA  95192-0106.}
\altaffiltext{8}{NASA Ames Research Center, MS 245-6, Moffett Field, CA  94035.}
\altaffiltext{9}{Department of Physics and Astronomy, Johns Hopkins University,
                 3400 North Charles Street, Baltimore, MD  21218.}
\altaffiltext{10}{I. Physikalisches Institut, Universit\"{a}t zu K\"{o}ln,
                 Z\"{u}lpicher Strasse 77, D-50937 K\"{o}ln, Germany.}

\begin{abstract}

We mapped a $13 \times 22$ pc region in emission from 492 GHz [\ci] and,
for the first time, 551 GHz \thcof\ in the giant molecular cloud M17SW.
The morphologies of the [\ci] and \thco\ emission are strikingly similar.
The extent and intensity of the [\ci] and \thcof\ emission is explained
as arising from photodissociation regions on the surfaces of embedded
molecular clumps.  Modeling of the \thcof\ emission in comparison to
\thcoo\ indicates a temperature gradient across the cloud, peaking to at
least 63 K near the M17 ionization front and decreasing to at least 20
K at the western edge of the cloud.  We see no correlation between gas
density and column density.  The beam-averaged column density of \ci\ in
the core is $1\times10^{18}$ \cmsqr, and the mean column density ratio
$N$(\ci)/$N$(CO) is about 0.4.  The variations of $N$(\ci)/$N$(CO) with
position in M17SW indicate a similar clump size distribution throughout
the cloud.

\end{abstract}

\keywords{ISM: individual (M17SW) --- ISM: molecules ---
ISM: structure --- radio lines: ISM --- submillimeter}

\section{INTRODUCTION}

M17SW is a giant molecular cloud at a distance of 2.2 kpc \citep{chi80}
illuminated by the M17 OB association about 1 pc to the east.
Studies of molecular excitation \citep{sne84}, molecular line profiles
\citep*{mar84}, and maps of molecular emission \citep{stu90} from the
M17SW core have all concluded that the dense gas is highly clumped.
Early photochemical models of molecular cloud/\hii\ region interfaces
predicted an overlying layer of ionized carbon and a thin underlying
layer of neutral carbon within a photodissociation region (PDR) between
the ionized and molecular gas components \citep*[e.g.,][]{lan76a,tie85}.
Given the edge-on geometry of M17SW and the \hii\ region, observations
of [\ci] and [C~{\sc ii}] emission along cuts through the interface region
unexpectedly detected these species more than a parsec into the
molecular cloud \citep{kee85,gen88,stu88}.  \citeauthor{gen88} and
\citeauthor{stu88} concluded that far-ultraviolet (FUV) radiation
penetrates the interclump medium and excites PDRs on the surfaces
of illuminated clumps.  Subsequent multi-component PDR models of
the cloud have accounted for the intensity of the emission from
numerous far-infrared and submillimeter atomic and molecular lines
\citep*{bur90,mei92}.  \citet{sek99} present observations of [\ci]
emission extending more than a degree southwest of the M17SW cloud core.

The dual radiometers on board the {\it Submillimeter Wave Astronomy
Satellite} (\swas) provide the ability to simultaneously observe the
emission from the $J = 5\rightarrow4$ rotational transition of \thco\ at
550.926 GHz and the $^3P_1\rightarrow{^3P_0}$ fine-structure transition of
\ci\ at 492.161 GHz.  The roughly 4\arcmin\ beamsize of \swas\ affords us
the opportunity to map large regions in the \thcof\ and [\ci] lines and
study the large-scale physical conditions in and structure of the dense
interstellar medium.  This is especially important for the \thcof\ line,
since few maps are available in the mid-$J$ \thco\ transitions beyond the
small-scale strip maps of \citet{gra93}.  Whereas the critical density
($n_{\rm cr} \sim 2\times10^3$ \cmcube) and upper-level energy ($E_u = 24$
K) of [\ci] enable it to be easily excited wherever neutral carbon is
present in the dense ISM, the \thcof\ line ($n_{\rm cr} \sim 2\times10^5$
\cmcube; $E_u = 79$ K) will preferentially probe warmer, denser regions.

We obtained with \swas\ the first large-scale map of the \thcof\ line
in M17SW, as well as the [\ci] line over the same region.  We combine
these results with millimeter \thco\ observations to probe the physical
conditions in the extended cloud and the spatial and density structure
of the bulk of the molecular gas.

\section{OBSERVATIONS AND RESULTS}

We mapped the [\ci] and \thcof\ lines in M17SW with \swas\ between 1999
June 30 and 1999 July 12.  We observed a $21\arcmin \times 35\arcmin$
region ($13 \times 22$ pc) in mapping mode with a 1\farcm6 grid
\hbox to\columnwidth{
spacing, nodding the telescope between the source position and
}

\centerline{\includegraphics[width=0.92\hsize,clip]{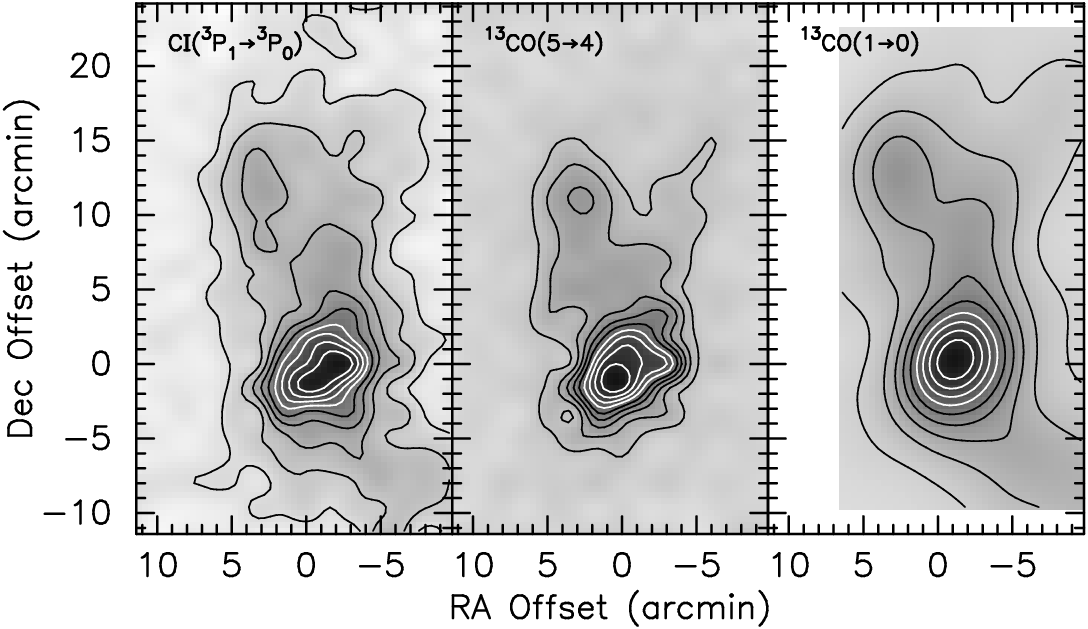}}
\figcaption{Integrated intensity maps of emission from 492
GHz [\ci] ({\it left panel}) and 551 GHz \thcof\ ({\it center panel}) in
M17SW observed by \swas, compared to a \thcoo\ map smoothed to equivalent
spatial resolution \citep[\it right panel\rm;][]{wil99}.  Mapping offsets
are relative to position $\alpha = 18^{\rm h}20^{\rm m}22\fs1$, $\delta =
-16\arcdeg12\arcmin37\arcsec$ (J2000).  Contour levels for all maps range
from 0.1 to 0.9 of the peak in steps of 0.1.  Peak intensities ($T_A^*$
scale) are 57.2 K \kms\ for [\ci], 56.9 K \kms\ for \thcof, and 38.2 K
\kms\ for \thcoo, integrated over the $V_{\rm LSR}$ range 14--26 \kms.
\label{fig1}}
\vspace{0.9\baselineskip}

\noindent
a reference position at $\alpha = 18^{\rm h}25^{\rm m}$13\fs1, $\delta =
-$16\arcdeg50\arcmin37\arcsec (J2000).  Integration times are typically
about 3 minutes, except for 16 central positions observed for H$_2$O
detections and described in \citet{sne00}.  The spatial resolution
of the observations is about 4\arcmin, and the spectral resolution
is about 1 \kms, sampled with a spacing of 0.6 \kms.  The [\ci] and
\thcof\ linewidths range from about 2--9 \kms\ (typically 5 \kms), so
the lines are well resolved.  We reduced the spectra using the standard
\swas\ pipeline procedure, and removed small first-order polynomial
baseline offsets.  Details of the \swas\ receivers, beam shapes, backend
spectrometer, and observing modes are presented in \citet{mel00}.
We present the maps herein on the $T_A^*$ scale, uncorrected for the
\swas\ main beam efficiency $\eta_{\rm mb} = 0.90$.

We show integrated intensity maps of the [\ci] and \thcof\ emission
in Figure~\ref{fig1}, as well as a \thcoo\ map from \citet*{wil99},
observed at 47\arcsec\ resolution with the 14 m telescope of the Five
College Radio Astronomy Observatory (FCRAO) but smoothed to 4\farcm0
resolution for comparison to the \swas\ observations.  The maps include
emission within the velocity range ($V_{\rm LSR}$) 14--26 \kms, which
excludes line-of-sight velocity components not associated with the
M17 \hii\ region.  The similarity of the morphology and extent of the
[\ci] and \thcoo\ emission is particularly striking.  This behavior of
the [\ci] and low-$J$ \thco\ emission has also been observed in other
giant molecular clouds, notably Orion A \citep{tau95,ike99,plu00}, S140
\citep*{plu94}, W3, L1630, and Cep A \citep{plu99}.  The morphology
of the \thcof\ emission is also similar to the [\ci] emission but
not as extended, except at the eastern edge nearest the \hii\ region.
The \thcof\ emission peaks slightly eastward of the [\ci] and \thcoo\
emission peaks, indicating higher temperatures or gas densities at the
eastern edge of the cloud.  We explore this further in \S3.

The shapes of the [\ci] and \thco\ lines are quite similar, with the
[\ci] and \thcoo\ line profiles being nearly identical.  There are some
subtle differences between the \thcof\ line and the other lines, most
notably a slight velocity shift at the map center position and slight
variances in the red side of the line profile in the northeastern region.
Overall, however, the spectra reveal a dynamical as well as morphological
similarity between the [\ci] and \thco\ gas components.

\centerline{\includegraphics[width=0.92\hsize,clip]{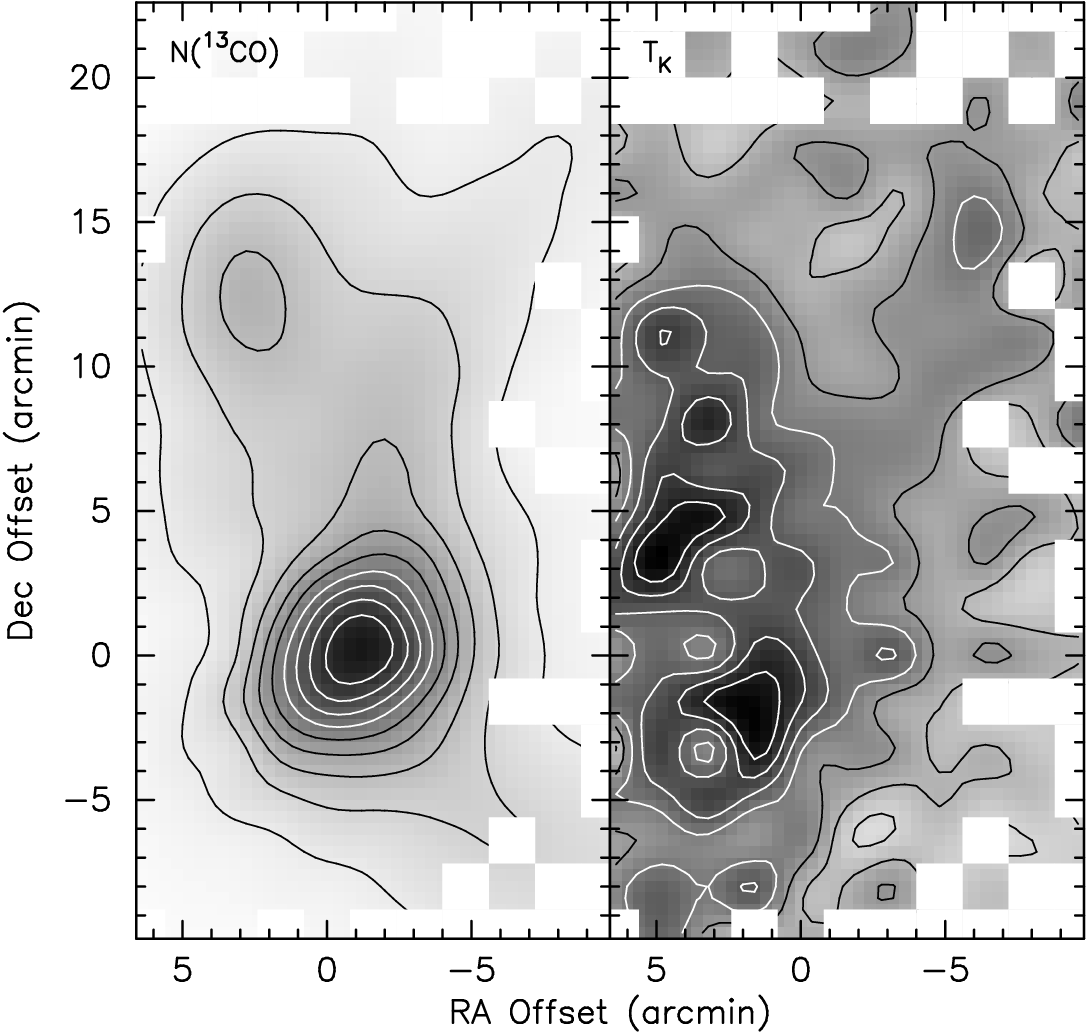}}
\figcaption{Model results for the distribution of \thco\
column density ({\it left panel}) and kinetic temperature ({\it
right panel}) in M17SW.  Position offsets are relative to $\alpha =
18^{\rm h}20^{\rm m}22\fs1$, $\delta = -16\arcdeg12\arcmin37\arcsec$
(J2000). Contour levels range from 0.1 to 0.9 of the peak in steps of 0.1
for $N$(\thco) (peak value $6.1\times10^{16}$ \cmsqr), and 0.3 to 0.9
of the peak in steps of 0.1 for $T_K$ (peak value 63~K).  Regions with
\thcof\ detections less than 1$\sigma$ are blanked.
\label{fig2}}

\section{DISCUSSION}

\subsection{Temperature, Density, and Column Density Structure}

We used the \thcof\ and \thcoo\ integrated intensities together with
an iterative algorithm incorporating the Large Velocity Gradient (LVG)
radiative transfer approximation \cite[e.g.,][]{sco74} to model the
\thco\ column density $N$(\thco), H$_2$ number density $n$(H$_2$),
and kinetic temperature $T_K$.  In the LVG modeling we used collision
rates for CO with para-H$_2$ \citep{flo85}.  Substituting ortho-H$_2$
rates resulted in insignificant differences in the model results.
For each position, we began with initial estimates of $T_K = 10$ K,
$N$(\thco) appropriate for LTE, and $n$(H$_2$) from \citet{sne84} for the
core or $n$(H$_2$) = $1 \times 10^5$ \cmcube\ elsewhere \citep{wil99}.
We used these estimates as input for the LVG algorithm, together with
velocity widths derived from the \thcoo\ data, to produce model \thcof\
and \thcoo\ integrated intensities.  We iterated the model with density
as the free parameter until the model \thcof/\thcoo\ integrated intensity
ratio reproduced the observations.  If the $J \leq 5$ level populations
thermalized with too low a ratio (i.e. further increases in density no
longer affected the line ratio), we incremented $T_K$ and reiterated
until the observed ratio was reached.  We then iterated the model with
$N$(\thco) as the free parameter until the absolute intensities of the
lines agreed with observations.  This value of $N$(\thco) was then used
again in the iteration with $n$(H$_2$) to check that line opacity effects
did not change the line ratio, and the entire loop repeated as necessary
until the model line intensities and ratios agreed with the observations.
This procedure then yields the minimum temperatures necessary to explain
the line ratios.

The model results show clear evidence that the molecular cloud
is heated by the \hii\ region/ionization front at its eastern edge.
In Figure~\ref{fig2} we show the temperature distribution over the
region mapped in \thcoo\ as well as the derived \thco\ column density
distribution.  The peak beam-averaged column
\hbox to\columnwidth{density is $6.1\times10^{16}$
\cmsqr, with a mean value of $1.2\times10^{16}$}

\centerline{\includegraphics[width=0.92\hsize,clip]{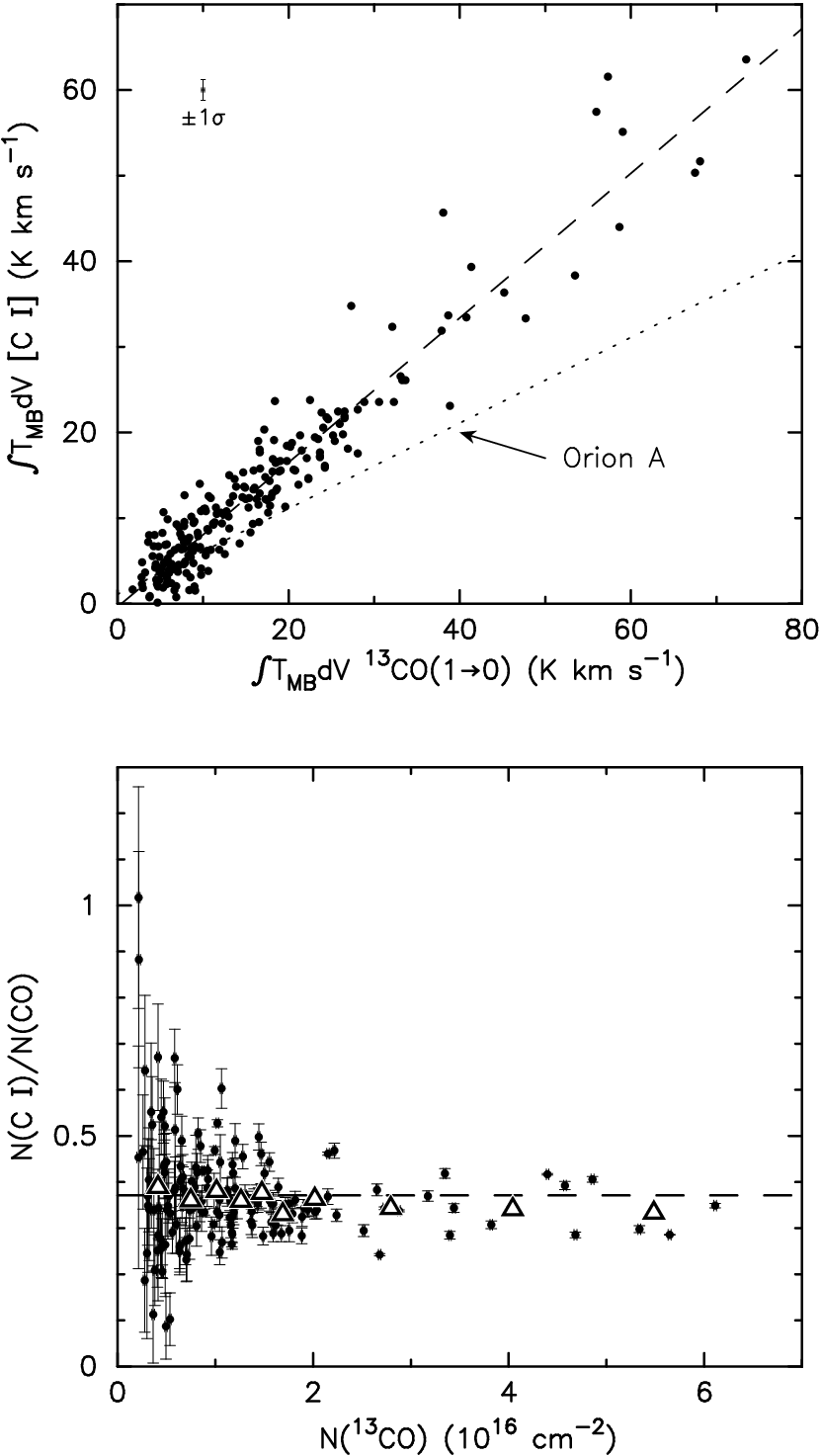}}
\figcaption{({\it Top}) The linear correlation of the
integrated intensity ($T_{\rm MB}$ scale) of [\ci] with \thcoo\ at each
position in M17SW.  The dashed line traces the least-squares linear
fit to the data (slope 0.84).  The dotted line plots the equivalent
fit for the Orion A cloud \citep[slope 0.5;][]{ike99}.  ({\it Bottom})
The derived ratio of the column densities of \ci\ to CO [where $N$(CO)
$\equiv 50N$(\thco)] is plotted against the \thco\ column density at
each position in M17SW.  Errorbars are $\pm$ 1$\sigma$ as described in
the text.  The dashed line marks the mean value of $N$(\ci)/$N$(CO) (0.37)
for all data points.  The triangles plot the value of $N$(\ci)/$N$(CO)
in bins containing equal numbers of \thco\ molecules, with the abscissa
at the center of each binning range.
\label{fig3}}
\centerline{ }
\centerline{ }

\noindent \cmsqr.  The column
density map appears virtually identical to the \thcoo\ integrated
intensity map (Fig.~\ref{fig1}), a property attributable to the low
opacity of the \thco\ lines ($\tau \lesssim 0.6$ for either line).  The
highest gas temperatures are found along the eastern edge of the cloud,
peaking at about 60~K near the cloud core.  The mean cloud temperature
is 35~K, similar to the global cloud temperature of 30~K derived by
\citet{wil99}.  The temperature we derive for the core region is also
consistent with the temperatures derived from  millimeter observations
of optically thin CH$_3$CCH emission \citep{ber94}.  Our observations
apparently are insensitive to the high temperature gas ($T_K \geq 200$
K) responsible for CO(7$\rightarrow$6) line emission observed in the
\hii/H$_2$ interface region, however this gas is estimated to comprise
only about 5 percent of the total column density \citep{har87} and would
contribute only about 10 percent of the observed peak \thcof\ intensity.

We derive molecular gas densities in the range 1--6 $\times 10^5$ \cmcube\
over the entire region mapped, with a median value of $3\times10^5$
\cmcube, about half as large as densities derived from multi-transition
CS observations of the core region of M17SW \citep{sne84,wan93}.
In contrast to the well-ordered column density and temperature
distribution across the M17SW cloud, the density distribution appears
almost random.  In particular, there is no correlation between the gas
density and the column density.  Instead, high- and low-density regions
are peppered randomly throughout the region.  Since the excitation
of the \thcof\ emission depends both on the temperature and density,
with higher densities requiring lower temperatures for the same $J =
5$ fractional population, we investigated whether there was any trend
between $T_K$ and $n$(H$_2$) as an artifact of the modeling algorithm.
We see no correlation between the gas temperature distribution and the
density distribution in the model results.

Although the model distributions of temperature and density are
uncorrelated, the densities we derive are sensitive to the model cloud
temperature.  For example, uniform cloud temperatures of 50~K and 100~K
yield median cloud densities of $4\times10^4$ and $1\times10^4$ \cmcube,
respectively.  The model column densities are insensitive to temperature,
decreasing by only 20 percent if $T_K = 100$~K.  Since such large column
densities of hot gas are difficult to produce throughout the cloud, and
since our model mean cloud temperature is already somewhat higher than
the \citet{wil99} global cloud temperature, we consider $4\times10^4$
\cmcube\ as a reasonable lower limit to the mean cloud density.

The modeling results for the density and column density of the molecular
gas in M17SW require that the gas is highly clumped.  We derive
the volume filling factor of the gas by dividing the ratio of the
peak H$_2$ column density ($4\times10^{22}$ \cmsqr, assuming a \thco\
abundance of $1.5\times10^{-6}$ relative to H$_2$) to the mean H$_2$
number density ($3\times10^5$ \cmcube) by the linear extent of the core
(3 pc core size, deconvolved from the \swas\ beam).  From these quantities,
we derive a volume filling factor of 0.02. \citet{stu90} derive
toward the central 13\arcsec\ of the core a volume filling factor of
0.13 for gas with densities greater than $10^5$ \cmcube, reflecting
a higher concentration of clumps at the core peak region.

\subsection{The Correlation of {\rm [C~{\sc i}]} and \thco\ Emission}

The large extent of the [\ci] emission in M17SW and its similarity to
isotopic CO emission was first noted by \citet{kee85} and \citet{gen88}
along linear cuts through the ionization front/molecular cloud interface.
The large-scale [\ci] and \thco\ maps in Figure~\ref{fig1} show that
this similarity extends over the entire region mapped.  We plot in
the top panel of Figure~\ref{fig3} the observed correlation of the
integrated intensities of the [\ci] and \thcoo\ emission.  We converted
the data to the main-beam temperature scale $T_{\rm mb}$ to remove
differences in telescope beam efficiencies from the observed relation
\citep[$\eta_{\rm mb} = 0.52$ for the FCRAO data;][]{wil99}.  The observed
trend is well fit (linear least-squares, correlation coefficient 0.95)
by the relation $\int T_{\rm mb}dV$([\ci])\ $ = (0.84\pm0.02)\int T_{\rm
mb}dV$(\thcoo) $-$ $(0.4\pm0.4)$.  The slope of this relation is similar
to the slope observed between [\ci] and \thcot\ in the Orion Bar region
by Tauber et al.\ (1995; $\sim 0.8$), but is somewhat higher than the
relation reported by Ikeda et al.\ (1999; $\sim 0.5$) for the Orion A
cloud as a whole, and for the ensemble of clouds W3, L1630, S140, and
Cep A \citep{plu99}.  We propose a possible explanation for this in \S3.3.

We used the \swas\ [\ci] observations to model the \ci\ column density
$N$(\ci) in a manner similar to the \thco\ LVG modeling, using the
collision rates of \citet{sch91}.  Since the spatial distribution and
velocity profiles of the \ci\ and \thco\ lines are nearly identical, we
used the temperatures and densities derived from the
\thco\ observations, along with the \thcoo\ linewidths, as fixed input
parameters to the LVG algorithm.  We then iterated the models with
$N$(\ci) as the sole free parameter until the model [\ci] intensities
matched the observations.  $N$(\ci) ranges from $1.1\times10^{16}$
\cmsqr\ to $1.1\times10^{18}$ \cmsqr\ with a median value of
$1.7\times10^{17}$ \cmsqr.  The [\ci] line is optically thin, with a
maximum opacity less than 0.6 at the core.  Our peak value of $N$(\ci)
is about half that obtained by \citet{sek99} in a 2\arcmin\ beam.
The spatial distribution of $N$(\ci) is nearly identical to the
distribution of $N$(\thco) shown in Figure~\ref{fig2}.

The derived \ci\ column densities show an interesting behavior when
compared to the \thco\ column densities.  In the bottom panel of
Figure~\ref{fig3} we plot the ratio of $N$(\ci)
to the CO column density $N$(CO) (calculated assuming a CO/\thco\
abundance ratio of 50) as a function of \thco\ column density for each
model position.  Errorbars reflect the random errors of the [\ci] and
\thcoo\ intensities, but not systematic uncertainties in the LVG
models.  Whereas at the low column densities at the edges of the
molecular cloud the $N$(\ci)/$N$(CO) ratio ranges from 0.1 to 1.0, at
high column densities in the core of the cloud the lower and upper
bound for the range asymptotically approach the mean $N$(\ci)/$N$(CO)
ratio for all positions (0.37).  This is similar to the behavior
exhibited for the Orion A cloud, except the mean column density ratio
is much lower, only about 0.10 in the Orion core regions
\citep{plu00}.  The behavior of the $N$(\ci)/$N$(CO) ratio is difficult
to understand in the context of chemical models where the neutral
carbon is produced {\it in situ} in the molecular gas
\citep*[e.g.,][]{lan76b,pin92}, and would thereby predict a more or
less constant column density ratio.  Likewise, simple plane parallel
PDR models \citep*[e.g.,][]{lan76a,tie85,hol91} predict a rising value
of $N$(\ci)/$N$(CO) with decreasing molecular gas column density, but
not the increased scatter we observe in the ratios.  Remarkably,
summing the $N$(\thco)-ordered data of Figure~\ref{fig3} into bins of
equal total number of \thco\ molecules (in this case, $2\times10^{17}$
\cmsqr\ $\times$ $A_{\rm beam}$, where $A_{\rm beam}$ is the area subtended by
the \swas\ beam at the distance of M17SW), and then computing
$N$(\ci)/$N$(CO) for the binned data, shows that $N$(\ci)/$N$(CO) is
independent of the mean \thco\ column density for each bin (see
Fig.~\ref{fig3}).  The resolution to the behavior of $N$(\ci)/$N$(CO)
lies in the structure of the M17SW molecular cloud.

\subsection{PDR Emission and the Structure of M17SW}

Theoretical models of PDRs are quite successful at predicting the
observed \thcof\ and [\ci] line intensities and the column density of \ci
for the physical conditions in the region mapped by \swas.  Based on the
arguments of \citet{stu88} and \citet{mei92}, we expect the incident FUV
field within the cloud to range from $\sim10$ $G_0$ at the northwestern
edge to $4\times10^4$ $G_0$ at the ionization front located about 1\farcm5
northeast of the map reference position, where $G_0$ is in units of the
Habing flux.  Our LVG modeling indicates a fairly narrow range of densities,
1--$6\times10^5$ \cmcube.  PDR models predict a column density $N$(\ci)
$\sim 4\times10^{17}$ \cmsqr\ insensitive to variations in $G_0$ and only
weakly dependent on density \citep{tie85}.  This is similar to the mean
value of $N$(\ci) derived for the M17SW cloud.  For the peak $N$(\ci)
we derive, only a few PDRs along the line of site through the core are
required.  \citet{kau99} present updated homogeneous PDR models covering
the range of densities and FUV fluxes appropriate for M17SW, and have
recently incorporated \thco\ into the models for comparison with \swas\
observations.  These models successfully predict the same range of
\thcof\ integrated intensities, for the densities and FUV fluxes we
find in M17SW, as those observed by \swas.  The PDR models of \citet{kos94}
also predict the peak \thcof\ line temperature we observe, for $G_0 = 10^4$
and $n$(H$_2$) $= 10^5$ \cmcube.  The \citet{kau99} models match
the observed [\ci] integrated intensities for regions of the cloud with
$G_0 < 100$, but are a factor of about 3--4 too low for the core of the
cloud.  Again, this is easily explained by requiring several PDRs along
the line of sight through the core.

The extent of the [\ci] emission and the relationship of $N$(\ci) with
the molecular gas column density require not only that the gas is
clumpy, but that the distribution of clump sizes is similar throughout
the cloud, from the core out to the periphery.  In a clumpy cloud
penetrated by FUV radiation, PDRs arise on the surfaces of the dense
clumps, with a thin shell of \ci\ encompassing the CO in the clump.
Consider unresolved clumps of similar density, as is the case for the
\swas\ observations \citep[see ][]{stu90}.  The beam-averaged column
density $N$(\ci) depends on the clump surface area while $N$(CO)
depends on the volume, so the ratio $N$(\ci)/$N$(CO) varies as the
inverse size of the clumps.  Thus, one might expect for a cloud with a
range of clump sizes a large scatter in the column density ratio, as we
see in the low column density regions where only a few clumps are
probably subtended by the \swas\ beam.  As the beam samples larger column
densities, i.e. more clumps in the beam, the clump size distribution is
more completely sampled and the scatter in $N$(\ci)/$N$(CO) decreases,
approaching the value appropriate for the mean surface area to volume
ratio of the clump size distribution.  By summing up the low column
density regions we increase the clump sample size, and we see from
Figure~\ref{fig3} that the $N$(\ci)/$N$(CO) ratio for equal total gas
masses gives the same value of $N$(\ci)/$N$(CO).  Thus, the size
distribution of clumps must be the same throughout the cloud.  The
diameter of a spherical clump with a \ci/CO number ratio of 0.37 and
H$_2$ density $3 \times 10^5$ \cmcube\ is 0.04 pc, where we assume all
carbon is either in \ci\ or in CO and calculate the thickness of the \ci
layer from \citet{hol91}.  This diameter scales inversely with the
H$_2$ density.  The lower $N$(\ci)/$N$(CO) ratio found in Orion is
probably a consequence of a larger mean clump size, since $N$(\ci) is
insensitive to changes in density and FUV flux.  Since the intensity of
optically thin lines scales directly with column density,
observationally a lower $N$(\ci)/$N$(CO) ratio produces a lower slope
in the linear correlation of the [\ci] line intensity to the
\thcoo\ intensity (e.g, Fig.~\ref{fig3}).

\acknowledgements

This work is supported by NASA contract NAS5-30702 and NSF grant AST 97-25951
to the Five College Radio Astronomy Observatory.  R. Schieder \& G. Winnewisser
would like to acknowledge the generous support provided by the DLR through
grants 50 0090 090 and 50 0099 011.

\end{document}